# Mid-infrared Energy Deposition Spectroscopy

Jiaze Yin[1,2], Christian Pfluegl[3], Chu C. Teng[3], Rylie Bolarinho[2,4], Guo Chen[1,2],

Xinrui Gong[2,5], Dashan Dong[1,2], Daryoosh Vakhshoori[3], Ji-Xin Cheng[1,2,4,5*]

1. Department of Electrical and Computer Engineering, Boston University, Boston, MA 02215, USA.

2. Photonics Center, Boston University, Boston, MA 02215, USA.

3. Pendar Technologies, Cambridge, MA 02138, USA

4. Department of Chemistry, Boston University, Boston, MA 02215, USA.

5. Department of Biomedical Engineering, Boston University, Boston, MA 02215, USA.

Correspondence to: jxcheng@bu.edu

**Abstract:**

Photothermal microscopy is an emerging tool for measuring light-matter interactions with single molecule sensitivity. It is generally believed that the spectral acquisition speed in photothermal microscopy is limited by the slow thermal diffusion process. Here, we demonstrate mid-infrared energy deposition (MIRED) spectroscopy that offers both microsecond-scale temporal resolution and sub-micron spatial resolution. In this approach, the photothermal process is optically probed while the infrared pulses from a quantum cascade laser array are rapidly tuned. Based on Newton's law, the energy deposition is the first derivative of local temperature rise over time and gives the instantaneous infrared absorption. By employing time-resolved measurement of transient energy deposition, the upper limit for spectrum encoding shifts to the vibrational relaxation level, which occurs on the picosecond scale. This method significantly increases the detection bandwidth while retaining the sensitivity and resolution benefits of photothermal detection.

By pump-probe measurement of absorption-induced change (temperature, size, refractive index, etc.) in a specimen, photothermal spectroscopy is an extremely sensitive analytical tool down to single molecule sensitivity[1,2]. Initial development of photothermal spectroscopy in the 1970s focused on the detection of weak overtone transitions in molecules[3]. Visible photothermal microscopy developed in the 1990s enabled label-free imaging of single gold nanoparticles down to a few nanometers in diameter[4]. In the early 2000s, the integration of atomic force microscopy (AFM) and infrared (IR) spectroscopy led to the development of AFM-IR that allows nanoscale examination of molecules in thin, dried specimens[5]. A more recent development of optical photothermal infrared spectroscopy, termed as mid-infrared photothermal (MIP) spectroscopy[6], has allowed high-resolution high-sensitivity infrared spectroscopic imaging of molecules in living cells at visible diffraction limit[7]. Several advanced methods have been developed, enabling scanning MIP imaging at video-rate[8,9], breaking the visible diffraction limits[10,11], high-throughput measurement in widefield manner[12-14], and volumetric chemical bond imaging through tomography[15,16].

Despite the extremely high sensitivity and resolution offered by photothermal detection, achieving rapid spectral acquisition via photothermal spectroscopy has been considered as a major obstacle. This limitation arises from the time needed for cooling during temperature modulation, which takes microseconds or longer. Consequently, even with an ultrafast wavelength-sweeping IR source and single pulse detection speed[8], the effective detection bandwidth is restricted to only hundreds of kilohertz per color. As shown in **Fig. 1(a)**, MIP spectroscopy uses a single-color IR source to excite a narrow band per measurement. The resulting temperature rise is linked to the absorption. Acquiring a spectrum requires continuously tuning the IR laser and inducing multiple cycles of photothermal modulation, with speeds typically around 0.1 seconds per 100 cm$^{-1}$ for a high-speed external cavity quantum cascade laser (EC-QCL).

In this work, we break the speed limitation by revisiting the photothermal process from the perspective of energy deposition. Based on Newton's law of heating and cooling[17] the temperature rise is the integration of heat generated during excitation as shown in Equation (1).

$$mC_s \frac{dT}{dt} = \dot{Q}_{abs} \qquad (1)$$

One can observe that the absorption coefficient $\alpha(\lambda)$ exists in the slope of the heating process instead of the overall amplitude by expand the heating term $\dot{Q}_{abs}$ as shown in Equation (2):

$$\dot{Q}_{abs} = I_{IR}(\lambda)\, \alpha(\lambda) \qquad (2)$$

Thus, by employing time-resolved measurement of transient energy deposition, the upper limit for spectrum encoding shifts to the vibrational relaxation level, which occurs on the picosecond scale[18,19]. This significant increase in detection bandwidth enables ultrafast IR spectroscopy while retaining the sensitivity benefits of photothermal detection.

To demonstrate the contrast mechanism of mid-infrared energy deposition (MIRED), we use a radiofrequency digitizer and measure the MIRED process induced by a broadband pulses train emission from a beam-combined monolithic distributed-feedback (DFB) QCL array[20,21]. Electric wavelength tuning of the laser array allows for broadband coverage with high resolution in the microsecond scale. This approach not only enhances spectrum acquisition speed into the microsecond scale but also enables the mapping of molecules with sub-micron resolution within complex biological samples. Technically, MIRED detection employs a wideband IR source that sequentially excites different molecular bands as shown in **Fig.1(b)**. Time-resolved measurement during the excitation process is performed. The absorption spectrum is derived from the transient energy deposition by taking the derivative of the heating curve over time. Spectrum acquisition finishes within the wideband pulse firing, which is on the microsecond scale. The theory of MIRED spectroscopy is detailed in **Supplementary Note 1.**

The setup for MIRED spectroscopy is illustrated in **Fig.S1 and Supplementary Method.** The QCL array consists of 32 independent channels, each set apart by 4 $cm^{-1}$, covering a range from 940 $cm^{-1}$ to 1056 $cm^{-1}$. In MIRED spectroscopy, each laser pulse is set to 100 ns wide, with all 32 channels firing sequentially within 3.2 μs. The characterization of the laser firing profile is shown in **Fig. S2**. The pulses train emission produces a wide band IR excitation. The central wavenumber of each laser is set around 4 $cm^{-1}$ apart with its neighbor. The central wavenumber together with power is listed as **Supplementary Table 1**. The induced heating profile of the absorber is probed by a continuous wave laser at 532 nm. The probe is detected by a fast photodiode and digitized at 4 ns per point, providing the necessary temporal resolution to resolve the energy deposition of each individual laser.

To validate the spectrum fidelity, MIRED spectra of three different chemicals were measured. The QCL array excites the sample from 940 $cm^{-1}$ to 1056 $cm^{-1}$ with all 32 channels firing sequentially. The power spectrum, shown in **Fig. 2(a)**, is used to normalize the absorption of each channel. The time-resolved heating curve under excitation is shown in **Fig. 2(b)**, with amplitude normalized to the same scale for comparison. The energy deposition is then evaluated by taking the derivative over time, as shown in **Fig. 2(c)**. With the system's temporal resolution at 4 ns, the energy deposition reveals the shape of each pulse in addition to the relative absorption. Subsequently, the time-domain energy deposition is converted into the spectra shown in **Fig. 2(d)**. In this window, the olive oil sample displays a strong peak at 966 cm$^{-1}$ from the -HC=CH- bending, dimethyl sulfoxide (DMSO) displays a strong absorption peak at 1022 cm$^{-1}$ from the S=O bond, while the glucose solution with a concentration of 200 mg/dL shows a broader peak around 1035 cm$^{-1}$ from the C-O bond. It is worth noting that, since the MIRED spectra are evaluated during a transient period with laser heating, the impact of thermal diffusion on the MIRED spectrum is negligible over such a short duration. The thermal diffusion effect is discussed in **Supplementary Note 2**.

To demonstrate the high temporal resolution advantage of MIRED spectroscopy, we performed continuous spectral acquisition at the rate of 20 microseconds per spectrum to visualize fast molecular solvation dynamics. The QCL array pulses train repeats at a frequency of 50 kHz, as shown in **Fig.S4**, providing a continuous spectrum output speed of 20 μs each. The binary mixtures of DMSO and water were used as the testing bed. DMSO and water mixtures exhibit interesting physical properties, such as dramatically changed freezing point and viscosity[22]. Solvation in these mixtures is known to be a highly dynamic process involving the breaking of hydrogen bonds and the formation of clusters[23]. Steady-state MIRED spectra of various concentrations of DMSO solution are shown in **Fig. 3(a)**. By reducing the percentage of DMSO, a redshift of the peak from 1022 cm$^{-1}$ to 1001 cm$^{-1}$ and decreased absorption can be observed. This can be explained by the decreased S=O dipole due to the formation of DMSO-water clusters while non-changed methyl group rocking[24]. The MIRED spectrum can sense such dipole changes as the intensity ratio between the two wavenumbers. We quantified the change in intensity ratio versus DMSO concentration, as shown in **Fig. 3(b)**. The blue curve represents the intensity ratio between the S=O peak at 1022 cm$^{-1}$ and the shift peak at 1001 cm$^{-1}$, while the red curve shows the intensity ratio between 1014 cm$^{-1}$ and 1001 cm$^{-1}$. The ratio is approximately 1.4 at 100% DMSO concentration and decreases to below 1 when the DMSO concentration is less than 10%, indicating a change in bond dipole in addition to the reduction in analyte concentration. For a low DMSO concentration solution, the water absorption overwhelms the MIRED intensity in all wavenumbers as shown in **Fig. 3(a)**, resulting in a base line background.

The molecular dynamics experiments begin with 100% DMSO, followed by the addition of excess water (20 times the volume of DMSO). The MIRED spectra are recorded continuously during the mixing process, as shown in **Fig. 3(c)**. With microsecond-scale temporal resolution, the MIRED amplitude at the specified wavenumbers can reveal the kinetics of cluster formation within 100 milliseconds. During this time, DMSO molecules undergo solvation at a rate of $k_S$ and cluster formation at a rate of $k_{CF}$ simultaneously. This cluster formation affects the dipole of the S=O bond, causing the peaks to shift. We model these two processes using linear kinetics, where at 1014 cm$^{-1}$, the intensity drop is governed by both processes with a combined lifetime.

$$\tau_{1014 \text{cm}^{-1}} = \frac{1}{k_S + k_{CF}} \qquad (3)$$

Conversely, the intensity at 1001 cm$^{-1}$, attributed to methyl rocking, is solely influenced by solvation. The lifetime is expressed as:

$$\tau_{1001 \text{cm}^{-1}} = \frac{1}{k_S} \qquad (4)$$

By fitting the MIRED intensity at 1014 cm$^{-1}$ and 1001 cm$^{-1}$, the lifetimes of the mixing kinetics are estimated to be 79 milliseconds and 103 milliseconds, respectively. Consequently, we can determine the cluster formation kinetics of water and DMSO molecules by solving Equation (3) and Equation (4). The results indicate that $k_{CF}$ is 2.95 s$^{-1}$, corresponding to a lifetime of 338 milliseconds.

To demonstrate MIRED spectroscopy on biological samples, we performed spectroscopic imaging of living cancer cells, as shown in **Fig. 4**. The MIRED spectra can differentiate between fatty acid esters and carbohydrates, enabling molecular mapping at submicron spatial resolution. During imaging, a piezo stage raster scans the sample while MIRED spectroscopy continuously records data. Spectra at each pixel are averaged 10 times, resulting in an effective pixel dwell time of 200 µs. The overall intensity, representing the total absorption of the sample, is shown in **Fig. 4(a)**. After converting each pixel into a spectrum, we can evaluate specific absorption bands of the sample, as shown in **Fig. 4(b)** and **(c)**. The merged image in **Fig. 4(d)** reveals the intracellular distribution difference between fatty acid esters and carbohydrates. Interestingly, aside from the diffuse carbohydrates spread throughout the cell body, a bright aggregate with 2.5 times stronger intensity than diffused carbohydrate is observed. This aggregate is distinct from lipid droplets, as indicated by the dashed line in **Fig. 4(e)**. These carbohydrate aggregates are likely glycogen in brain cancer cells[25,26]. The average spectra of the two molecule types are shown in **Fig. 4(f)**, illustrating spatial spectrum heterogeneity. To further validate the MIRED content mapping of biomolecules, the cancer cells cultured at delipidated (charcoal-striped) serum, in which most of the hydrophobic lipid species were removed, were measured. Quantification analysis showed that the lipid depletion group had the dropped intensity in lipid channel from 966 $cm^{-1}$ to 981 $cm^{-1}$ compared with control group as shown in **Fig. S5**, while there is no significant change on carbohydrate channel from 1022 $cm^{-1}$ to 1057 $cm^{-1}$, suggesting the accuracy of applying MIRED for study molecules inside cells. In addition to imaging living cells, we demonstrate MIRED spectroscopic imaging of large tissue samples at submicron resolution. MIRED imaging provides hyperspectral data with a single scan of the sample, significantly reducing the acquisition time. The mouse brain imaging results are shown in **Fig. S6**, with pseudo colors red, green, and blue representing the average absorption from windows of 980-1010 cm$^{-1}$, 1018-1031 cm$^{-1}$, and 1035-1057 cm$^{-1}$, respectively. In the MIRED image, structures such as the cerebrum, corpus callosum, striatum, and are resolved [27] based on their chemical composition differences. The cerebrum is uniformly enriched with carbohydrates where the fatty acid ester channel (980-1010 cm$^{-1}$) is darker than the carbohydrate channel (1035-1057 cm$^{-1}$), indicating a high population of neurons and high carbohydrate levels in the brain. The corpus callosum is primarily composed of fiber bundles, showing strong absorption similar to fat. In between, the cerebral cortex region demonstrates a transition between the fat signal in white matter (inside the brain) and the relatively high sugar signal in gray matter.

Vibrational spectroscopy for chemical detection has greatly advanced in two major directions: increasing the speed to detect molecular processes in highly dynamic environments[28-32] and enhancing the sensitivity to visualize small amounts of substances in complex systems. In this work, we have developed MIRED-based vibrational spectroscopy and imaging methods that bridge these two directions by measuring energy deposition via a pump-probe scheme. MIRED enables single-shot spectral acquisition on a microsecond time scale and imaging with submicron spatial resolution. This method opens a new way of using infrared spectroscopy to perform temporally and spatially resolved chemical analysis in various fields of material science and life science.

By revisiting the energy deposition process during laser excitation, we found that the absorption coefficient can be accurately determined via time-resolved pump-probe measurement of the heating slope. Given the vibrational relaxation time at the picosecond scale[18,19], fast spectral encoding and decoding can be implemented using a wideband infrared pulse and RF detection without cross-talk. This represents a dramatic improvement in bandwidth for spectrum information transfer. In the current setup employing a QCL array, the spectrum range is limited by the number of excitation sources. This can be further improved by combining arrays covering different IR emission bands, such as 128 channels covering the entire fingerprint region[33]. Additionally, the concept of MIRED can be extended to measuring thermal effects induced by sources with ultrafast sweeping in nanoseconds-sale, such as mid-infrared frequency combs[34] and time-stretched laser pulses[35,36], for further speed improvement.

Notably, MIRED spectroscopy offers solution-phase molecular detection where most chemistry and biology occur. In such environments, conventional IR loss measurements require strictly controlled sample thickness and solvent attenuation to collect enough infrared photons on the detector[37]. Using the pump-probe method, MIRED relaxes these conditions by measuring the transient effects occurring precisely at the absorption site. The extension of ultrafast vibrational spectroscopy in this domain would significantly benefit biological research, aiding in the understanding of chemical reactions involved in cell metabolism and the conformation of biomolecules at the subcellular level. For example, the recently demonstrated use of molecular probes to sense enzymatic reactions via the nitrile group in the cell's silent window[38] could benefit from MIRED. This technique would enhance quantitative analysis and background correction by enabling the acquisition of the entire spectrum. Additionally, MIRED spectroscopy holds the potential for high-speed mid-infrared flow cytometry, thanks to its spectral throughput for screening cells for specific biomolecular markers with high sensitivity.

It is noteworthy that the speed of vibrational spectroscopy is ultimately constrained by the number of measurable absorption events occurring upon excitation. MIRED spectroscopy inherently offers higher sensitivity in measuring these events for two main reasons. Firstly, it doesn't measure the intense excitation field, resulting in a relatively low background compared to direct photon loss measurement[7]. Secondly, sensitivity is enhanced by eliminating the need for mid-infrared photon detectors, which typically suffer from thermal noise and low quantum efficiency at high bandwidth. Collectively, transient energy deposition measurement holds the potential for ultrafast vibrational spectroscopy in analysis, where both sensitivity and speed are crucial.

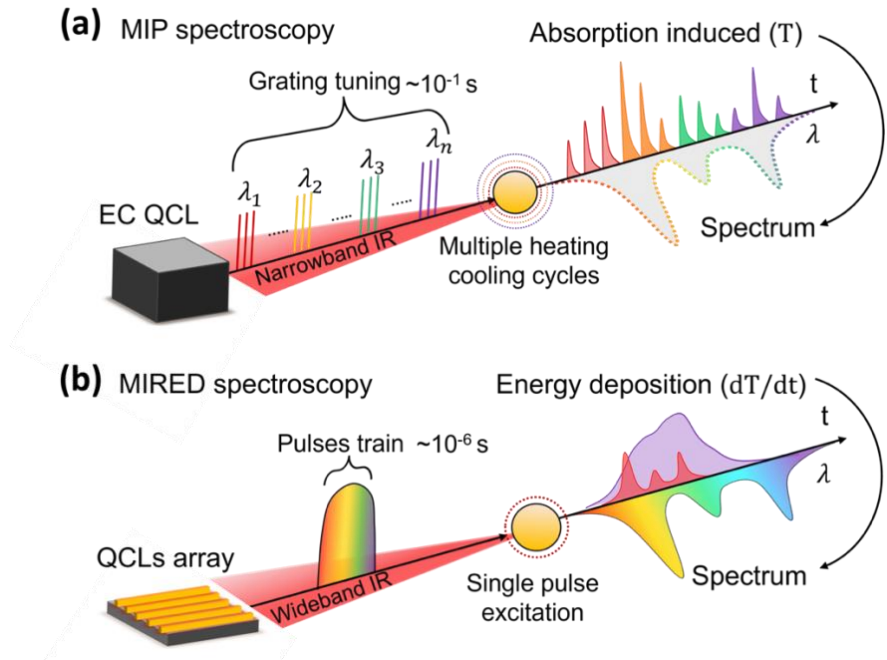

**Figure 1. Comparison of MIP spectroscopy using a QCL and MIRED spectroscopy using a broadband excitation source.** (a) MIP signal is from a single-color excitation source. Measuring an absorption spectrum requires tuning through discrete IR wavelengths and involves multiple cycles of heating and cooling. (b) MIRED spectroscopy using a broadband excitation source. By performing time-resolved energy deposition measurement in nanosecond scale, an absorption spectrum is obtained from the first derivative of the energy deposition trace.

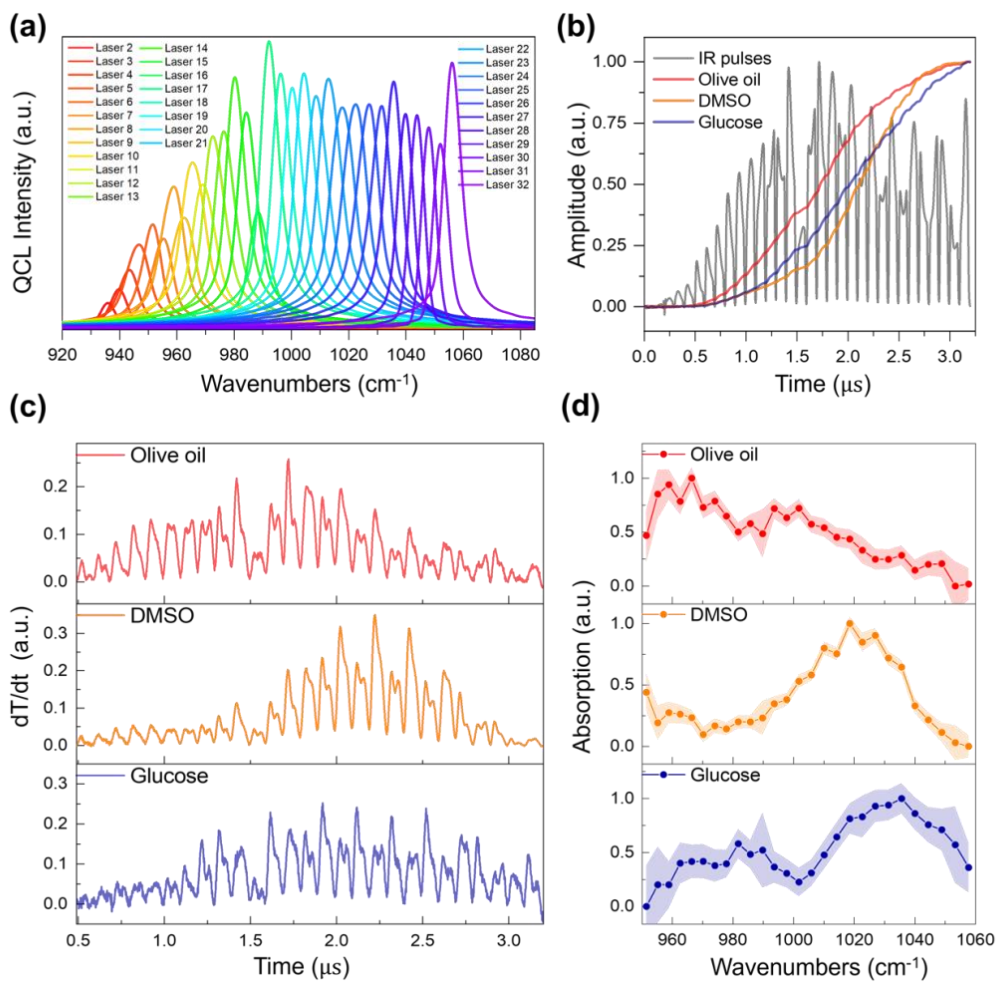

**Figure 2. MIRED spectra of olive oil, dimethyl sulfoxide (DMSO), and glucose solution.** (a) Characterized power spectrum of QCL array. (b) Pulses train fired by QCLs array (gray) and induced heating curve of olive oil (red), DMSO (orange), and glucose solution with 200mg/dL (blue). (c) Time-resolved energy deposition of the corresponding chemical in (b). (d) Resolved MIRED spectra from the time-domain after normalizing with pulse energy. Twenty measurements are performed and averaging results are displayed. In panel d, the data are presented as the mean ± standard deviations.

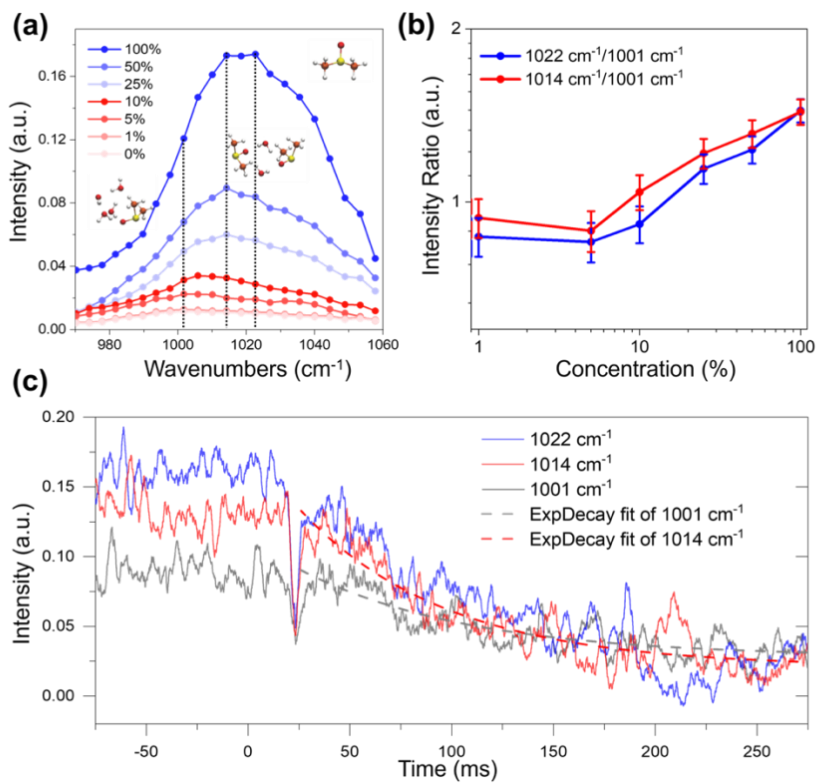

**Figure 3. MIRED spectroscopy of solvation dynamics on the microsecond scale.** (a) MIRED spectrum of DMSO-water binary mixture at different DMSO ratios. Instead of decreased absorption, a red-shift is observed for the DMSO peak as the DMSO concentration decreased. (b) MIRED intensity ratio between two wavenumbers versus DMSO concentration. Fifty measurements were performed, and the averaged results are displayed. The data are presented as the mean ± standard deviation. (c) Molecular solvation dynamics during the DMSO water mixing process revealed by single pulse MIRED spectroscopy. 200 μL DI water is added to 10 μL DMSO by pipette at t=0. The MIRED intensity at the indicated wavenumbers decreases over time at different ratios. The exponential decay is fit on the MIRED intensity at 1014 cm$^{-1}$ and 1001 cm$^{-1}$, the time constants are 79 milliseconds and 103 milliseconds, respectively.

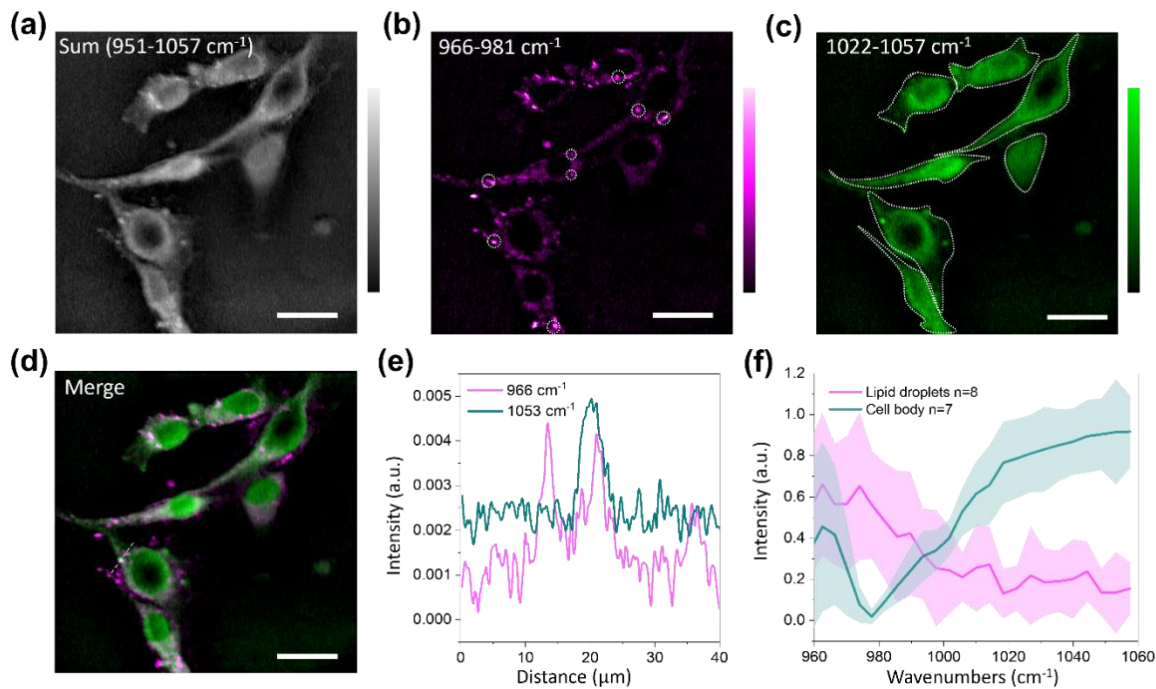

**Figure 4. MIRED spectroscopic imaging of live U87 cancer cells.** (a) Average intensity from 951 cm$^{-1}$ to 1057 cm$^{-1}$. (b) Average intensity from 966 cm$^{-1}$ to 981 cm$^{-1}$, the lipid droplets show strong absorption. (c) Average intensity from 1022 cm$^{-1}$ to 1057 cm$^{-1}$, the carbohydrate contributes to the absorption. (d) Merge image of lipid ester (Magenta), and carbohydrate (Green). (e) Line profile of indicated area in merge image (d) reveals the carbohydrate aggregation. (f) The MIRED spectra of lipid droplets and cell body. N=8, 6 accordingly for statistical analysis. The data are presented as the mean ± standard deviations.

**Acknowledgements:** This work is supported NIH grants R35GM136223, R33CA261726, R01 EB032391, R01 EB035429 to JXC.

**Author contributions:** J.Y. developed the MIRED spectroscopy system, designed the experiments, processed the data and drafted the paper. C.P., C.T., and D.V. provided the QCL array and technical support. R.B. prepared cancer cell samples. G.C. and X.G. provided the mouse tissue samples. D.D. provided the FT-IR measurement. J.-X.C. initialized the project, revised the paper and provided scientific guidance.

**Competing Interests:** C.P., C.C., D.V. work at a company "Pendar Technologies" that incorporates the QCL into commercial products. J.-X.C. declares financial interest with Photothermal Spectroscopy Corp, which did not support this work. The other authors declare no competing interests.

**Data availability:** All data needed to evaluate the conclusions in the paper are present in the paper and/or the Supplementary Materials. Raw data and code are available in Zenodo (DOI: 10.5281/zenodo.12741205).

**References**

[1] A. Gaiduk, M. Yorulmaz, P. Ruijgrok, and M. Orrit, Science **330**, 353 (2010).

[2] S. Adhikari, P. Spaeth, A. Kar, M. D. Baaske, S. Khatua, and M. Orrit, ACS nano **14**, 16414 (2020).

[3] M. Long, R. L. Swofford, and A. Albrecht, Science **191**, 183 (1976).

[4] S. Berciaud, L. Cognet, G. A. Blab, and B. Lounis, Physical review letters **93**, 257402 (2004).

[5] A. Dazzi and C. B. Prater, Chem. Rev. **117**, 5146 (2017).

[6] D. Zhang, C. Li, C. Zhang, M. N. Slipchenko, G. Eakins, and J.-X. Cheng, Science advances **2**, e1600521 (2016).

[7] Y. Bai, J. Yin, and J.-X. Cheng, Science Advances **7**, eabg1559 (2021).

[8] J. Yin, M. Zhang, Y. Tan, Z. Guo, H. He, L. Lan, and J.-X. Cheng, Science Advances **9**, eadg8814 (2023).

[9] G. Ishigane, K. Toda, M. Tamamitsu, H. Shimada, V. R. Badarla, and T. Ideguchi, Light: Science & Applications **12**, 174 (2023).

[10] P. Fu, W. Cao, T. Chen, X. Huang, T. Le, S. Zhu, D.-W. Wang, H. J. Lee, and D. Zhang, Nature Photonics, 1 (2023).

[11] M. Tamamitsu, K. Toda, M. Fukushima, V. R. Badarla, H. Shimada, S. Ota, K. Konishi, and T. Ideguchi, Nature Photonics, 1 (2024).

[12] Y. Bai, D. Zhang, L. Lan, Y. Huang, K. Maize, A. Shakouri, and J.-X. Cheng, Science advances **5**, eaav7127 (2019).

[13] K. Kniazev, E. Zaitsev, S. Zhang, Y. Ding, L. Ngo, Z. Zhang, G. V. Hartland, and M. Kuno, ACS Photonics **10**, 2854 (2023).

[14]	T. Yuan, L. Riobo, F. Gasparin, V. Ntziachristos, and M. A. Pleitez, Science Advances **10**, eadj7944 (2024).

[15]	J. Zhao *et al.*, Nature Communications **13**, 7767 (2022).

[16]	M. Tamamitsu, K. Toda, H. Shimada, T. Honda, M. Takarada, K. Okabe, Y. Nagashima, R. Horisaki, and T. Ideguchi, Optica **7**, 359 (2020).

[17]	T. L. Bergman, F. P. Incropera, D. P. DeWitt, and A. S. Lavine, *Fundamentals of heat and mass transfer* (John Wiley & Sons, 2011).

[18]	X. Chen, Y. Chen, M. Yan, and M. Qiu, ACS nano **6**, 2550 (2012).

[19]	J. C. Lindon, G. E. Tranter, and D. Koppenaal, *Encyclopedia of spectroscopy and spectrometry* (Academic Press, 2016).

[20]	B. G. Lee *et al.*, IEEE Journal of Quantum Electronics **45**, 554 (2009).

[21]	M. F. Witinski *et al.*, Optics Express **26**, 12159 (2018).

[22]	R. N. Havemeyer, Journal of pharmaceutical sciences **55**, 851 (1966).

[23]	A. Idrissi, B. Marekha, M. Kiselev, and P. Jedlovszky, Physical Chemistry Chemical Physics **17**, 3470 (2015).

[24]	V. M. Wallace, N. R. Dhumal, F. M. Zehentbauer, H. J. Kim, and J. Kiefer, The Journal of Physical Chemistry B **119**, 14780 (2015).

[25]	E. Favaro *et al.*, Cell metabolism **16**, 751 (2012).

[26]	J. Pelletier, G. Bellot, P. Gounon, S. Lacas-Gervais, J. Pouysségur, and N. M. Mazure, Frontiers in oncology **2**, 18 (2012).

[27]	G. A. Johnson *et al.*, Proceedings of the National Academy of Sciences **120**, e2218617120 (2023).

[28]	J. Réhault, F. Crisafi, V. Kumar, G. Ciardi, M. Marangoni, G. Cerullo, and D. Polli, Optics express **23**, 25235 (2015).

[29]     M. Tamamitsu, Y. Sakaki, T. Nakamura, G. K. Podagatlapalli, T. Ideguchi, and K. Goda, Vibrational Spectroscopy **91**, 163 (2017).

[30]     C.-S. Liao, P. Wang, P. Wang, J. Li, H. J. Lee, G. Eakins, and J.-X. Cheng, Science Advances **1**, e1500738 (2015).

[31]     K. Goda and B. Jalali, Nature Photonics **7**, 102 (2013).

[32]     N. Picqué and T. W. Hänsch, Nature Photonics **13**, 146 (2019).

[33]     C.-S. Liao, R. Blanchard, C. Pfluegl, M. Azimi, F. Huettig, and D. Vakhshoori, Optics Letters **45**, 3248 (2020).

[34]     D. A. Long, M. J. Cich, C. Mathurin, A. T. Heiniger, G. C. Mathews, A. Frymire, and G. B. Rieker, Nature Photonics **18**, 127 (2024).

[35]     A. Kawai, K. Hashimoto, T. Dougakiuchi, V. R. Badarla, T. Imamura, T. Edamura, and T. Ideguchi, Communications Physics **3**, 152 (2020).

[36]     K. Hashimoto, T. Nakamura, T. Kageyama, V. R. Badarla, H. Shimada, R. Horisaki, and T. Ideguchi, Light: Science & Applications **12**, 48 (2023).

[37]     I. Pupeza *et al.*, Nature **577**, 52 (2020).

[38]     H. He *et al.*, Nature methods, 1 (2024).